\documentclass[groupedaddress,twocolumn,prd,preprintnumbers,amsmath,amssymb,nofootinbib]{revtex4}

\usepackage{graphicx}
\usepackage{dcolumn}
\usepackage{bm}
\usepackage{epsfig}
\usepackage{epstopdf}
\usepackage{amssymb}
\usepackage{amsmath}
\usepackage{subfigure}
\usepackage{color}
\usepackage[dvipsnames]{xcolor}
\usepackage[utf8]{inputenc}

\newcommand{\beq}{\begin{equation}}
\newcommand{\eeq}{\end{equation}}

\newcommand{\eq}{\mathrm{eq}}

\newcommand{\sigunits}{\,\mathrm{cm^3\,s^{-1}}}

\begin{document}

\title{Growth of Dark Matter Perturbations during Kination}

\author{Kayla Redmond}
\email{kayla.jaye@unc.edu}
\author{Anthony Trezza}
\author{Adrienne L. Erickcek}
\email{erickcek@physics.unc.edu}
\affiliation{Department of Physics and Astronomy, University of North Carolina at Chapel Hill, Phillips Hall CB3255, Chapel Hill, NC 27599 USA}

\begin{abstract}
If the Universe's energy density was dominated by a fast-rolling scalar field while the radiation bath was hot enough to thermally produce dark matter, then dark matter with larger-than-canonical annihilation cross sections can generate the observed dark matter relic abundance.  To further constrain these scenarios, we investigate the evolution of small-scale density perturbations during such a period of kination.  We determine that once a perturbation mode enters the horizon during kination, the gravitational potential drops sharply and begins to oscillate and decay.  Nevertheless, dark matter density perturbations that enter the horizon during an era of kination grow linearly with the scale factor prior to the onset of radiation domination.  Consequently, kination leaves a distinctive imprint on the matter power spectrum: scales that enter the horizon during kination have enhanced inhomogeneity.  We also consider how matter density perturbations evolve when the dominant component of the Universe has a generic equation-of-state parameter $w$.  We find that matter density perturbations do not grow if they enter the horizon when ${0< w < 1/3}$.  If matter density perturbations enter the horizon when ${w > 1/3}$, their growth is faster than the logarithmic growth experienced during radiation domination.  The resulting boost to the small-scale matter power spectrum leads to the formation of enhanced substructure, which effectively increases the dark matter annihilation rate and could make thermal dark matter production during an era of kination incompatible with observations.
\end{abstract}

\maketitle

%%%%%%%%%%%%%%%
\section{Introduction}
\label{sec:Intro}
%%%%%%%%%%%%%%%%

There are no direct observational probes of the period between the end of inflation and the beginning of Big Bang Nucleosynthesis (BBN), and as a result, our understanding of this period is severely limited.  Unfortunately, this ignorance hinders our ability to understand both baryogenesis and dark matter production \cite[e.g.][]{Giudice:2000Paper}.  There is hope that the spectrum of gravitational waves generated prior to BBN could probe this era, but these probes either require futuristic gravitational wave detectors \cite{Boyle:2005GW, Easther:2006GW, Easther:2008GW, Amin:2014, Giblin:2014GW} or a network of cosmic strings \cite{Cui:2017GW}.  The matter power spectrum provides another way to probe this era.  For example, an early-matter-dominated era (EMDE) prior to BBN enhances the small-scale matter power spectrum and increases the abundance of microhalos \cite{Erickcek:2011,Erickcek:2015}.  These microhalos enhance the dark matter annihilation rate by several orders of magnitude, depending on the cutoff in the small-scale matter power spectrum.  These boosted annihilation rates are sufficient to bring some EMDE scenarios with otherwise undetectable dark matter particles into tension with Fermi-LAT observations of dwarf spheroidal galaxies \cite{Erickcek:2015,Erickcek:Sinha:Watson:2015}.

Another possibility is that there was a period of kination between the end of inflation and the beginning of BBN, during which the Universe was dominated by a fast-rolling scalar field (a kinaton) \cite{Spokoiny:1993, Joyce:1996, Ferreira:1997}.  Kination was initially proposed as a post-inflationary model that allows the Universe to transition to radiation domination even if the inflaton does not fully decay into radiation \cite{Spokoiny:1993}.  Kination also facilitates baryogenesis \cite{Joyce:1996}, and the kinaton can mimic the effects of a cosmological constant if its potential energy becomes dominant at very late times \cite{Ferreira:1997, Peebles:1998, Dimopoulos:2001, Dimopoulos:2002Curvaton, Chung:2007}.

In Ref.~\cite{Redmond:2017}, we explored how the dark matter density evolves if it is thermally produced during an era of kination, and we derived analytic expressions for the dark matter relic abundance; see also Refs.~\cite{Profumo:2003, Pallis:2005, Pallis:2nd2005, Pallis:2006, Gomez:2008, Lola:2009, Pallis:2009, DEramo:2017, Visinelli:2017, DEramo:2017ecx}.  To obtain the observed dark matter relic abundance, dark matter that is thermally produced during an era of kination requires larger annihilation cross sections than dark matter that is thermally produced during radiation domination.  Using recent observational limits on dark matter annihilations within dwarf spheroidal galaxies \cite{Fermi:Constraints} and the Galactic Center \cite{HESS:Constraints}, we were able to place tight constraints on the dark matter mass and the temperature at which kinaton-radiation equality occurs, provided that the dark matter reaches thermal equilibrium during an era of kination \cite{Redmond:2017}.

In this work, we study what effect kination has on the growth of dark matter density perturbations. If kination enhances the growth of dark matter density perturbations, the resulting small-scale structure would increase the dark matter annihilation rate.  This boost to the annihilation rate would place even tighter constraints on scenarios where dark matter reaches thermal equilibrium during an era of kination.  If the growth of perturbations during kination amplifies the dark matter annihilation rate by a factor of 10, then dark matter that is thermally produced during kination and annihilates via the $b\overline{b}$, $\tau^+ \tau^-$, or $W^+W^-$ annihilation channels will be ruled out \cite{Redmond:2017}.

First, we numerically determine the evolution of cosmological perturbations during an era of kination.  Surprisingly, we find that dark matter density perturbations grow linearly with the scale factor for perturbation modes that enter the horizon during kination.  To better understand this linear growth, we derive analytic expressions for the evolution of the gravitational potential $\Phi$ and fractional dark matter density perturbation $\delta_\chi$, not only during an era of kination, but also for scenarios where the dominant component of the Universe has a generic equation-of-state parameter $w$.  We determine that once a mode enters the horizon, the gravitational potential drops sharply and then oscillates with a decaying amplitude if the dominant energy density has ${w > 0}$.  In addition, if ${w > 1/3}$, then ${\delta_\chi \propto a^{3w/2 - 1/2}}$, where $a$ is the scale factor.  Therefore, if a perturbation mode enters the horizon during an era of kination $(w = 1)$, then $\delta_\chi$ grows linearly with the scale factor.  This growth leaves an imprint on the matter power spectrum.  We determine that for modes that enter the horizon during an era of kination, ${\delta_\chi/\Phi_0 \propto k^{1/2}}$, where k is the comoving wave number, and $\Phi_0$ is the value of the gravitational potential on superhorizon scales during kination.

Our perturbation analysis is applicable for scenarios in which dark matter does and does not reach thermal equilibrium during an era of kination.  References~\cite{Redmond:2017, DEramo:2017} determined that if dark matter reaches thermal equilibrium during an era of kination, annihilations do not cease until after the Universe becomes radiation dominated.  We determine that these ``relentless" annihilations do not significantly influence the evolution of $\delta_\chi$ after a mode has entered the horizon.  Since dark matter annihilation cannot lead to deviations from adiabaticity on superhorizon scales \cite{Weinberg:2003,Weinberg:2004,Weinberg:2004Second}, ``relentless" annihilation has a minimal effect on the matter power spectrum.

In Section \ref{sec:Perturbation}, we present the evolution equations that govern density and velocity perturbations.  In Sections \ref{sec:PhiEvolution} and \ref{sec:ChiEvolution}, we derive analytic expressions for the evolution of the gravitational potential and dark matter density perturbations, respectively.   In Section \ref{sec:MatterPowerSpectrum}, we determine how the matter power spectrum scales with wave number following an era of kination.  In Section \ref{sec:Conclusion}, we summarize our results and discuss their implications.  The appendices detail the derivation of the perturbation evolution equations and their initial conditions.  Natural units $(\hbar = c=k_B=1)$ are used throughout this work.

%%%%%%%%%%%%%%%
\section{Perturbation Evolution}
\label{sec:Perturbation}
%%%%%%%%%%%%%%%

We consider a three-fluid model consisting of dark matter, radiation, and the kinaton.  The kinaton is a fast-rolling scalar field: ${w \equiv P_\phi/\rho_\phi \simeq 1}$, where $P_\phi$ is the kinaton pressure and $\rho_\phi$ is the kinaton energy density.  We assume that the dark matter is composed of Majorana particles and that the kinaton does not decay nor otherwise interact with radiation or dark matter.  However, dark matter and radiation are thermally coupled via pair production and annihilation.  Therefore, the equations for $\rho_\phi$, the radiation energy density $\rho_r$, and the dark matter number density $n_\chi$ are
\begin{subequations}
\begin{align}
& \frac{d}{dt} \rho_{\phi} = -6H\rho_{\phi}, \\
& \frac{d}{dt}n_{\chi} = -3Hn_{\chi} - \langle \sigma v \rangle (n_{\chi}^2 - n_{\chi,\eq}^2), \\
& \frac{d}{dt}\rho_r = -4H\rho_r + \langle \sigma v \rangle E_{\chi} (n_{\chi}^2 - n_{\chi,\eq}^2),
\end{align}
\label{eq:Boltz}%
\end{subequations}
where $\langle \sigma v \rangle$ is the velocity-averaged dark matter annihilation cross section, ${ \langle E_{\chi} \rangle = \rho_{\chi}/n_{\chi}}$ is the average energy of a dark matter particle, and $n_{\chi,\eq}$ is the number density of dark matter particles in thermal equilibrium.  For a dark matter particle with mass $m_{\chi}$ and internal degrees of freedom $g_{\chi}$ within a thermal bath of temperature $T$,
\begin{align}
n_{\chi,\eq} = \frac{g_{\chi}}{2\pi^2} \int_{m_{\chi}}^{\infty} \frac{\sqrt{E^2 - m_{\chi}^2}}{e^{E/T} +1} E \, \mathrm{d}E .
\label{eq:Equilibrium}
\end{align}
We approximate $n_{\chi,\eq}$ as
\begin{align}
n_{\chi,\eq} \simeq \frac{g_{\chi}}{2\pi^2} \, m_{\chi}^2 \, T \, K_2\left(\frac{m_\chi}{T}\right),
\label{eq:EquilibriumBessel}
\end{align}
where $K_2(z)$ is a modified Bessel function of the second kind.  Equation~(\ref{eq:EquilibriumBessel}) matches Eq.~(\ref{eq:Equilibrium}) to within $0.1\%$ for $m_\chi/T \gtrsim 6$.  In addition, when evaluating ${ \langle E_{\chi} \rangle}$, we make the approximation that ${\langle E_{\chi} \rangle \simeq \sqrt{m_{\chi}^2 + (3.151 \, T)^2}}$, which matches ${\rho_{\chi}/n_{\chi}}$ to within $10\%$.

Since the kinaton does not interact with either the radiation or dark matter, the kinaton energy density scales as ${\rho_\phi \propto a^{-6}}$.  Even though the dark matter and radiation are thermally coupled via pair production and annihilation, this interaction is not sufficient to influence the evolution of the radiation energy density $\rho_r$, and thus, ${\rho_r \propto a^{-4}}$ \cite{Redmond:2017}.  If dark matter does not reach thermal equilibrium, the dark matter ``freezes in".  In these scenarios, after pair production ceases (when $T \lesssim m_\chi/4$), the dark matter energy density scales as ${\rho_\chi \propto a^{-3}}$.  If dark matter does reach thermal equilibrium, the dark matter ``freezes out", and annihilations do not cease until after kinaton-radiation equality \cite{Redmond:2017,DEramo:2017}.  As a result, ${\rho_\chi \propto (a^3 [1+ \, \mathrm{ln}(a/a_\mathrm{f})])^{-1}}$ between freeze-out ${(a = a_\mathrm{f})}$ and kinaton-radiation equality \cite{Redmond:2017}.\footnote{The logarithmic scaling of ${\rho_\chi a^3}$ is the same for kination models and cannibalistic dark matter models.  In cannibalistic dark matter models, the dark matter undergoes self-heating, which produces a pressure term in the perturbation equations \cite{Carlson:1992Cann, Machacek:1994Cann, deLaix:1995Cann, Buen-Abad:2018Cann}.  This pressure term is absent in kination models because the energy released by dark matter annihilations is transferred to the radiation bath.}

Perturbation modes are characterized by their comoving wave number $k$.  A perturbation mode enters the horizon when ${k = aH}$, where $H$ is the Hubble parameter.  Each fluid has fractional density perturbations ${\delta_i \equiv (\rho_i - \rho^0_i)/\rho^0_i}$, where $\rho^0_i(t)$ is the fluid's background energy density.  Each fluid also has velocity perturbations ${\theta_i \equiv a\partial_j v^j}$, where ${v^j = \mathrm{d}x^j/\mathrm{d}t}$ is the fluid's comoving peculiar velocity.  We assume that the relativistic particles are tightly coupled so that we may neglect the higher moments of the radiation perturbation.

In Appendix \ref{sec:perts} we present the derivation of the perturbation evolution equations using the conformal Newtonian gauge.  These equations govern the evolution of ${\delta}$ and $\theta$ for all fluids.  To numerically evaluate the perturbation equations, we rewrite them in terms of the scale factor and dimensionless parameters.  We define ${E(a) \equiv H(a)/H_1}$, ${\tilde{k} \equiv k/H_1}$, and ${\tilde{\theta}_i\equiv\theta_i /H_1}$, where ${H_1 \equiv H(a=1)}$ and ${a=1}$ is the start of the numerical integration.  Using these conventions, the perturbation equations for the kinaton $\phi$, radiation $r$, and dark matter $\chi$ are
\begin{widetext}
\begin{subequations}
\begin{align}
& \delta_\phi'+\frac{2\tilde{\theta}_\phi}{a^2 E(a)} + 6\Phi' = 0,\\
& \tilde{\theta}'_\phi - 2\frac{\tilde{\theta}_\phi}{a} +\frac{\tilde{k}^2 \Phi}{a^2 E(a)} - \frac{1}{2}\frac{\tilde{k}^2 \delta_\phi}{a^2 E(a)}=0,\\
& \delta'_\chi+\frac{\tilde{\theta}_\chi}{a^2 E(a)} + 3\Phi' = \frac{\langle \sigma v \rangle \rho^0_\chi}{m_\chi H_1 a E(a)} \left[\Phi\left\{1-\left(\frac{\rho^0_{\chi,\mathrm{eq}}}{\rho^0_\chi} \right)^2\right\} - \delta_\chi + \left(\frac{\rho^0_{\chi,\mathrm{eq}}}{\rho^0_\chi}\right)^2 \left(2\delta_{\chi,\mathrm{eq}}-\delta_\chi \right) \right],\\
& \tilde{\theta}'_\chi +\frac{\tilde{\theta}_\chi}{a} +\frac{\tilde{k}^2 \Phi}{a^2 E(a)} = \frac{\langle \sigma v \rangle \left(\rho^0_{\chi,\mathrm{eq}}\right)^2}{m_\chi \rho^0_\chi H_1 a E(a)} \left(\tilde{\theta}_r - \tilde{\theta}_\chi \right),\\
& \delta'_r+\frac{4}{3}\frac{\tilde{\theta}_r}{a^2 E(a)} + 4\Phi' =  \frac{\langle \sigma v \rangle \left(\rho^0_\chi\right)^2}{m_\chi \rho^0_r H_1 a E(a)} \left[-\Phi\left\{1-\left(\frac{\rho^0_{\chi,\mathrm{eq}}}{\rho^0_\chi} \right)^2\right\} + 2\delta_\chi - \delta_r - \left(\frac{\rho^0_{\chi,\mathrm{eq}}}{\rho^0_\chi}\right)^2 \left(2\delta_{\chi,\mathrm{eq}}-\delta_r \right) \right],\\
& \tilde{\theta}'_r + \frac{\tilde{k}^2 \Phi}{a^2 E(a)} - \frac{1}{4}\frac{\tilde{k}^2 \delta_r}{a^2 E(a)}= \frac{\langle \sigma v \rangle \left(\rho^0_\chi\right)^2}{m_\chi \rho^0_r H_1 a E(a)} \left(\frac{3}{4} \tilde{\theta}_\chi - \tilde{\theta}_r + \frac{1}{4} \left(\frac{\rho^0_{\chi,\mathrm{eq}}}{\rho^0_\chi} \right)^2\tilde{\theta}_r   \right),
\end{align}
\label{pertsa}%
\end{subequations}
\end{widetext}
\begin{figure*}[t]
 \centering
\begin{minipage}{0.5\textwidth}
\centering
 \resizebox{3.4in}{!}
 {
      \includegraphics{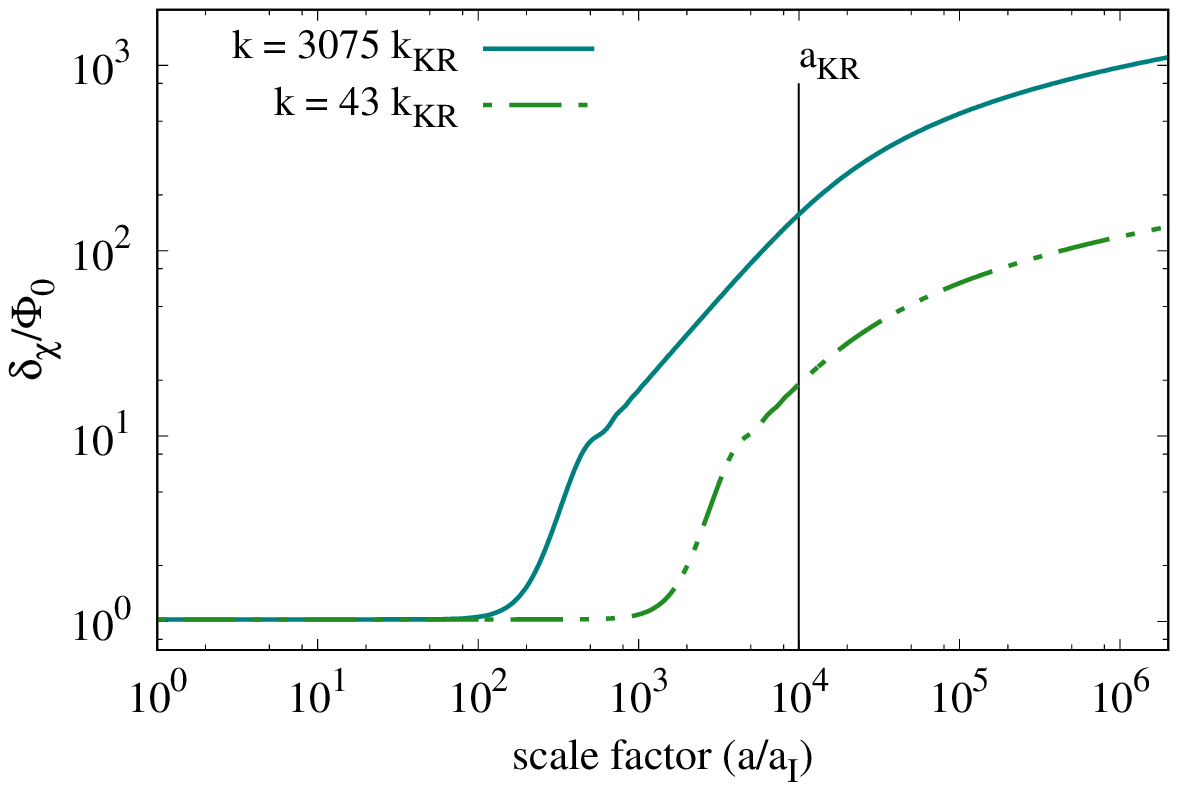}
 }
\end{minipage}%
\begin{minipage}{0.5\textwidth}
\centering
 \resizebox{3.4in}{!}
{
      \includegraphics{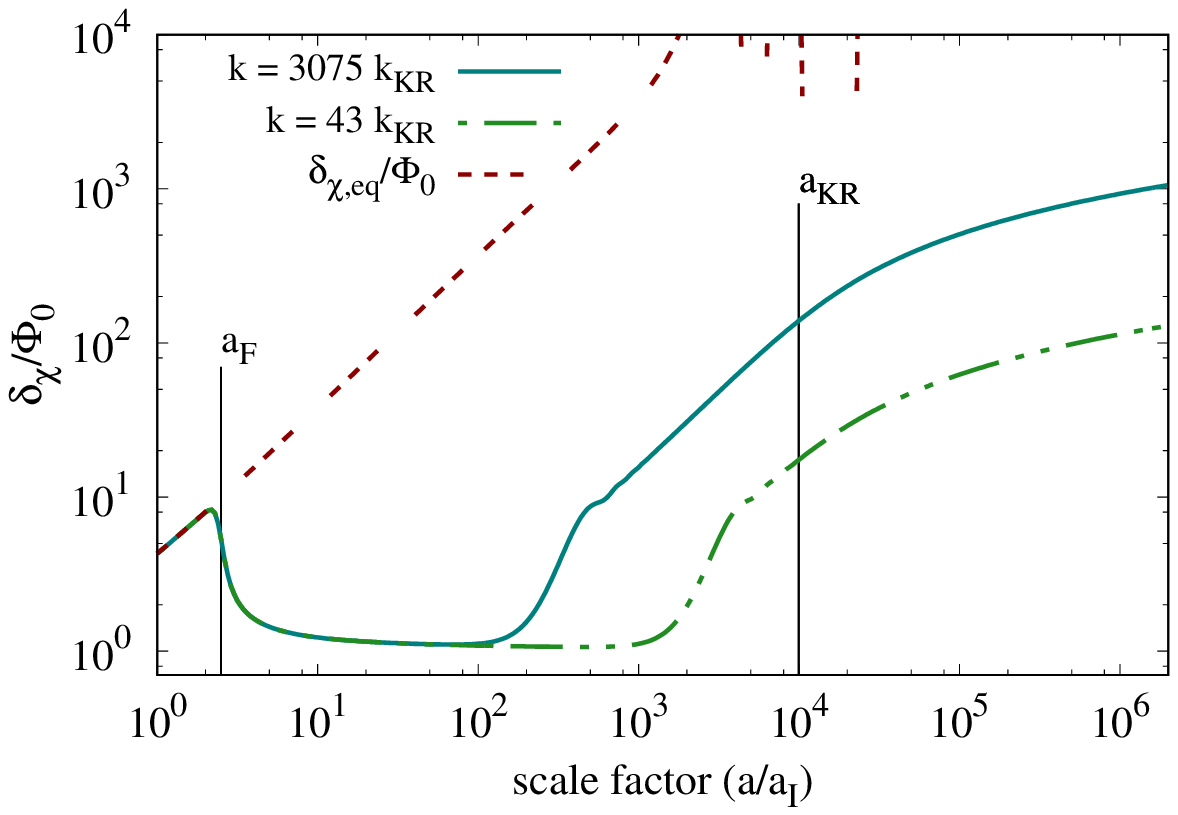}
 }
\end{minipage}%
\caption{The evolution of dark matter density perturbations for two modes that enter the horizon during an era of kination.  The left panel corresponds to scenarios where dark matter freezes in during an era of kination with ${\langle \sigma v \rangle = 1.4 \times 10^{-46} \sigunits}$.  The right panel corresponds to scenarios where dark matter freezes out during an era of kination with ${\langle \sigma v \rangle = 1 \times 10^{-23} \sigunits}$ and freeze-out occurring at ${a_\mathrm{F}=2.5}$.  The short-dashed line corresponds to the dark matter equilibrium density perturbation ${\delta_{\chi,\mathrm{eq}}}$.  In both panels, ${m_\chi = 10^5 \, \mathrm{GeV}}$ and kinaton-radiation equality occurs at ${a_\mathrm{KR} = 10^4}$.  One mode has wave number ${k = 3075\, k_{\mathrm{KR}}}$ and enters the horizon at ${a = 150}$, while the other mode has wave number ${k = 43\, k_{\mathrm{KR}}}$ and enters the horizon at ${a = 1300}$.}
\label{Fig:ChiEvolution}
\end{figure*}
where $'$ denotes a derivative with respect to the scale factor, ${\delta_{\chi,\mathrm{eq}} \equiv n_{\chi,\mathrm{eq}}/n^0_{\chi,\mathrm{eq}}-1}$ is the dark matter equilibrium density perturbation, and $\Phi$ and $\Psi$ are metric perturbations [see Eq.~(\ref{eq:PerturbedFRW})].  In addition, the perturbed time-time component of the Einstein equation yields
\begin{align}
\tilde{k}^2 \Phi + 3a^{2}E^2(a)\left(\Phi' a + \Phi \right) = \frac{3}{2} a^{2}\left(\tilde{\rho}_\phi \delta_\phi + \tilde{\rho}_r \delta_r + \tilde{\rho}_\chi \delta_\chi \right),
\label{eq:PerturbedEinstein}
\end{align}
where ${\tilde{\rho_i}\equiv \rho^0_i/\rho_c}$ and ${\rho_c \equiv 3H^2_1 \, m^2_{\mathrm{PL}}/8\pi}$.  When deriving Eqs.~(\ref{pertsa}) and (\ref{eq:PerturbedEinstein}), we use the fact that, since scalar fields cannot have anisotropic stress to first order in the perturbations, $\Phi = -\Psi$.

We numerically solve Eq.~(\ref{pertsa}) for various $k$ values starting well before each mode enters the horizon and after the dark matter becomes nonrelativistic ${(m_\chi/T \gtrsim 3)}$.  For any given $k$ mode, we assume that the perturbations are adiabatic before horizon entry.\footnote{References \cite{Weinberg:2004,Weinberg:2004Second} demonstrated that perturbations that are initially adiabatic remain adiabatic before horizon entry even in the presence of energy exchange between fluids.}  This implies that the initial perturbations are all directly related to the initial gravitational potential $\Phi_0$: see Appendix \ref{sec:InitialConditions}.

It has been well established that for modes that enter the horizon during radiation domination, matter density perturbations grow logarithically with the scale factor \cite{Hu:1995}.  Once the Universe becomes matter dominated, the matter density perturbations grow linearly with the scale factor.  By numerically solving the perturbation evolution equations, we determine that if a mode enters the horizon during an era of kination, matter density perturbations grow linearly with the scale factor prior to the onset of radiation domination.  Figure \ref{Fig:ChiEvolution} shows the evolution of dark matter density perturbations obtained by numerically solving Eq.~(\ref{pertsa}) for freeze-in and freeze-out scenarios.  The two modes shown in Figure \ref{Fig:ChiEvolution} both enter the horizon during an era of kination.  The two modes have wave numbers ${k = 43 \, k_{\mathrm{KR}}}$ and ${k = 3075 \, k_{\mathrm{KR}}}$, where $k_{\mathrm{KR}} \equiv a_{\mathrm{KR}} H(a_{\mathrm{KR}})$ is the wave number of the mode that enters the horizon at kinaton-radiation equality.  The two modes enter the horizon respectively at ${a = 1300}$ and ${a = 150}$, while kinaton-radiation equality occurs at ${a_\mathrm{KR} = 10^4}$.  In the freeze-in scenarios, the dark matter density perturbations are initially constant in conformal Newtonian gauge.  Since our perturbation evolution starts after pair production has mostly ended, ${\rho_\chi \propto a^{-3}}$.  To ensure that the curvature perturbation ${\zeta_\chi = \Phi- \delta_\chi \rho_\chi / (a \rho_\chi')}$ remains constant outside the horizon, $\delta_\chi$ must also be constant outside the horizon for freeze-in scenarios.  The situation is more complicated for freeze-out scenarios because ${\delta_\chi = \delta_{\chi,\mathrm{eq}}}$ until freeze-out, after which $\delta_\chi$ decreases toward $\Phi_0$ to maintain adiabaticity before horizon entry.

Once a mode enters the horizon, the dark matter density perturbation experiences a kick from the decaying gravitational potential (see Figure \ref{Fig:PhiEvolutionAnalytic}).  After the kick, the dark matter density perturbations grow linearly with the scale factor until kinaton-radiation equality, after which they grow logarithmically.  The evolution of dark matter density perturbations is oddly similar during an era of kination and matter domination, in spite of the fact that the pressure of the kinaton forces $\Phi$ to evolve very differently during an era of kination.  In the following sections, we analytically solve for the evolution of $\Phi$ and $\delta_\chi$ in order to determine the physical mechanism behind the linear growth of $\delta_\chi$ during kination.

%%%%%%%%%%%%%%%
\section{Analytic Expressions}
\label{sec:Analytics}
%%%%%%%%%%%%%%%

%%%%%%%%%%%%%%%
\subsection{$\Phi$ Evolution}
\label{sec:PhiEvolution}
%%%%%%%%%%%%%%%

To understand the evolution of $\delta_\chi$, we must first understand the evolution of $\Phi$.  To do so, we compare how $\Phi$ evolves for modes that enter the horizon during various eras.  To form a single differential equation for $\Phi$, we start with the time-time and space-space components of the perturbed Einstein equations:
\begin{subequations}
\begin{align}
k^2 \Phi + 3\frac{\dot{a}}{a} \left(\dot{\Phi} - \frac{\dot{a}}{a}\Psi \right) &= 4\pi G a^2 \delta\rho, \\
\ddot{\Phi} + \frac{\dot{a}}{a} \left(2 \dot{\Phi} - \dot{\Psi}  \right) - \left( 2 \frac{\ddot{a}}{a} - \frac{\dot{a}^2}{a^2} \right)\Psi \\
+ \frac{k^2}{3}\left(\Phi + \Psi \right) &=  -4\pi Ga^{2} \delta P, \nonumber
\end{align}
\label{eq:EinsteinPerturbationsW}%
\end{subequations}
where a dot represents differentiation with respect to conformal time, and $\delta\rho$ and $\delta P$ are the dominant fluid's energy density and pressure perturbations.  Assuming that the dominant fluid has a constant equation of state, ${\delta P = w \,\delta\rho}$.  Combining Eq.~(\ref{eq:EinsteinPerturbationsW}) with the second Friedmann equation yields a second-order differential equation for $\Phi$ that is dependent on $w$:
\begin{align}
\ddot{\Phi} + \frac{6(1+w)}{3w+1}\tau^{-1}\dot{\Phi} + wk^2\Phi = 0,
\label{eq:PhiDifferentialW}
\end{align}
where $\tau$ is the conformal time.  The solution to Eq.~(\ref{eq:PhiDifferentialW}) for ${w>0}$ is
\begin{align}
\Phi(\tau) = C_1 \, &\left(2 \tau + 6 w \tau \right)^b \, J_{-b}\left(w^{1/2} k \, \tau \right) \nonumber \\
& + C_2 \, \left(2 \tau + 6 w \tau \right)^b \, Y_{-b}\left(w^{1/2} k \, \tau \right),
\label{eq:ConformalPhiW}
\end{align}
where $b = 1/2- 3(1+w)/(3w+1)$, $C_1$ and $C_2$ are integration constants, $J_b$ is a Bessel function of the first kind, and $Y_b$ is a Bessel function of the second kind.  Conformal time and the scale factor are related via $w$: since ${H(a) = H_1 a^{-3(1+w)/2}}$, ${\tau = [H_1(3w+1)/2]^{-1}\,a^{(3w+1)/2} }$.  Using this relation, Eq.~(\ref{eq:ConformalPhiW}) is rewritten in terms of the scale factor, and $C_1$ and $C_2$ are determined by demanding that $\Phi \rightarrow \Phi_0$ as $a \rightarrow 0$:
\begin{align}
\Phi(a) = \Phi_0 \, \Gamma(1-b) \, &w^{\frac{b}{2}} \left(\frac{\tilde{k}}{3w+1} \right)^{b} a^{\frac{b}{2} (3w+1)} \nonumber \\
& \times J_{-b}\left(w^{1/2} \frac{2\tilde{k}}{3w+1} a^\frac{3w+1}{2} \right),
\label{eq:PhiAnalyticAFullW}
\end{align}
where $\Gamma(x)$ is the Gamma function.

\begin{figure}
\centering\includegraphics[width=3.4in]{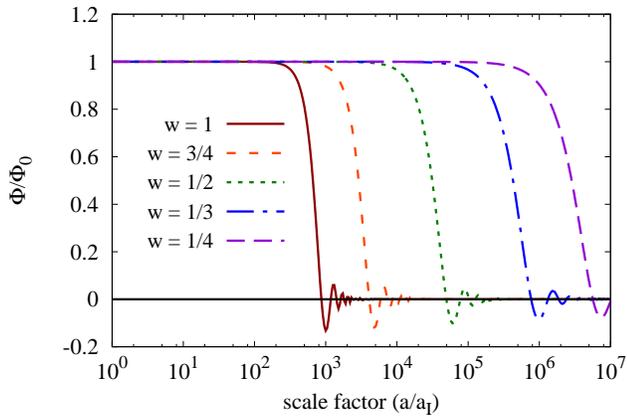}
\caption{The evolution of $\Phi$ given that the perturbation mode enters the horizon when the dominant component of the Universe has ${w = 1, 0.75, 0.5, 0.33, \mathrm{or} \, \, 0.25}$.  All of these scenarios involve a mode with ${k/H_1 = 10^{-5}}$, but they enter the horizon at varying values of $a$.}
\label{Fig:PhiEvolutionAnalyticGenericW}
\end{figure}

Figure \ref{Fig:PhiEvolutionAnalyticGenericW} shows the analytic evolution of $\Phi$ given by Eq.~(\ref{eq:PhiAnalyticAFullW}).  Each curve represents a perturbation mode that enters the horizon when the dominant component of the Universe has ${w = 1, 0.75, 0.5, 0.33, \mathrm{or} \, \, 0.25}$.  All the curves have ${\tilde{k} = 10^{-5}}$, and therefore the ratio between $k$ and $H_1$ is the same for all scenarios, and the perturbation mode in each scenario starts out equally far outside the horizon.  However, since ${E(a) \propto a^{-3(1+w)/2}}$, these modes enter the horizon at different values of $a$.  Apart from the varying horizon entries, the overall evolution of $\Phi$ is qualitatively the same for ${w>0}$.  Once a perturbation mode enters the horizon, the fluid's pressure overwhelms the gravitational attraction, causing a sudden drop in $\Phi$, after which $\Phi$ oscillates with a decaying amplitude.  Setting ${w>1}$, which does not necessarily violate causality \cite{ArmendarizPicon:1999, Christopherson:2008, DEramo:2017}, leads to the same $\Phi$ evolution.  In contrast, $\Phi$ is constant during matter domination.  Therefore, the linear growth experienced by matter perturbations that enter the horizon during an era of kination is fundamentally different from the linear growth experienced during matter domination.

\begin{figure}
\centering\includegraphics[width=3.4in]{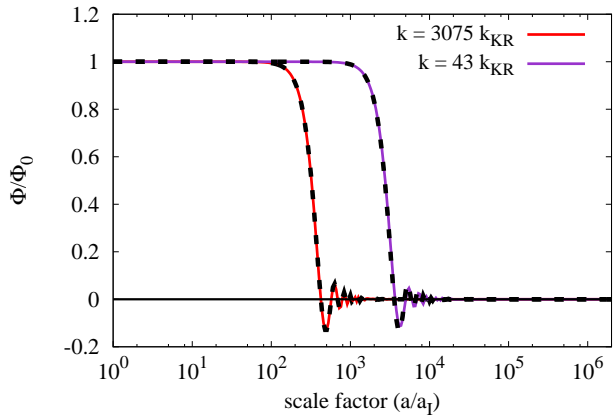}
\caption{The evolution of $\Phi$ for two modes that enter the horizon during an era of kination.  In this figure, ${m_\chi = 10^5 \, \mathrm{GeV}}$, ${\langle \sigma v \rangle = 1.4 \times 10^{-46} \sigunits}$, and kinaton-radiation equality occurs at ${a_\mathrm{KR}= 10^4}$.  One mode has wave number ${k = 3075 \, k_{\mathrm{KR}}}$ and enters the horizon at ${a = 150}$, while the other mode has wave number ${k = 43 \, k_{\mathrm{KR}}}$ and enters the horizon at ${a = 1300}$.  The solid lines represent the numeric solution to Eq.~(\ref{eq:PerturbedEinstein}) and the dashed lines represent the analytic approximations for $\Phi$ using Eq.~(\ref{eq:PhiKination}).}
\label{Fig:PhiEvolutionAnalytic}
\end{figure}

During an era of kination, ${w=1}$, and
\begin{align}
\Phi(a) = \Phi_0 \, \frac{4}{\tilde{k} a^2} J_1 \left(\frac{\tilde{k}a^2}{2} \right),
\label{eq:PhiKination}
\end{align}
since $\Gamma(2)=1$.  Expanding the Bessel function in Eq.~(\ref{eq:PhiKination}) around ${\tilde{k}a^2 = 0}$ reproduces the initial condition for $\Phi$ derived in Appendix \ref{sec:InitialConditions}: ${\Phi \simeq \Phi_0 - \tilde{k}^2 \Phi_0 \, a^4/32}$ for $\tilde{k}^2 a^4 \ll 1$.  Figure \ref{Fig:PhiEvolutionAnalytic} shows the evolution of $\Phi$ for the same modes as those shown in Figure \ref{Fig:ChiEvolution} (the evolution of $\Phi$ is identical for freeze-in and freeze-out scenarios since ${\rho_\chi \ll \rho_\phi}$).  The solid curves represent the numeric solution to Eq.~(\ref{eq:PerturbedEinstein}) and the dashed lines represent the analytic approximation given by Eq.~(\ref{eq:PhiKination}).  We see from Figure \ref{Fig:PhiEvolutionAnalytic} that $\Phi$ remains constant while the perturbation mode is outside the horizon.  Upon horizon entry, $\Phi$ drops sharply and begins to oscillate around zero with a decaying amplitude.  The percent error between the numeric and analytic solutions for $\Phi$ remains below $0.1\%$ as long as $a \lesssim 0.25 \, a_{\mathrm{KR}}$.  For $a \lesssim 0.25 \, a_{\mathrm{KR}}$, the assumption that $\rho_\phi$ is the sole component of the Universe is valid because $\rho_\phi$ contributes at least $95\%$ of the total energy density of the Universe.  If a perturbation mode enters the horizon near kinaton-radiation equality, Eq.~(\ref{eq:PhiKination}) becomes inaccurate because it assumes that the kinaton is the sole component of the Universe.

%%%%%%%%%%%%%%%
\subsection{$\delta_\chi$ Evolution}
\label{sec:ChiEvolution}
%%%%%%%%%%%%%%%

If a perturbation mode enters the horizon during an era of kination, we have seen numerically that the dark matter density perturbation experiences linear growth.  By deriving an analytic expression for $\delta_\chi$ we will gain an understanding of where this linear growth originates.  In the limit of kinaton domination, Eqs.~(\ref{pertsa}c) and (\ref{pertsa}d) can be rewritten as
\begin{subequations}
\begin{align}
& \delta'_\chi = -\tilde{\theta}_\chi a - 3\Phi', \\
& \tilde{\theta}'_\chi = -\frac{\tilde{\theta}_\chi}{a} - a\tilde{k}^2 \Phi.
\end{align}
\label{eq:KinationLimitPerturbationsDM}%
\end{subequations}
From these equations, we derive a single second order differential equation for $\delta_\chi$:
\begin{align}
\delta^{\prime \prime}_\chi  = \Phi\left(\frac{a^2}{a^4_\mathrm{hor}} \right) - 3\Phi^{\prime \prime} \equiv S(k,a),
\label{eq:DMSecondOrder}
\end{align}
where $a_\mathrm{hor}$ is the scale factor value at horizon entry.  The right hand side of Eq.~(\ref{eq:DMSecondOrder}) is the source term $S(k,a)$.  To solve Eq.~(\ref{eq:DMSecondOrder}), we first solve the homogeneous equation ${\delta^{\prime \prime}_\chi = 0}$: $\delta_\chi = C_1 + C_2 \, a$, where $C_1$ and $C_2$ are integration constants.  The full solution to Eq.~(\ref{eq:DMSecondOrder}) is a combination of the homogeneous solution and the particular solution.  The particular solution is the integral of the source term times the Green's function (GF).  The Green's function itself is a linear combination of the homogeneous solutions.  If the homogeneous solutions are $D_1(a)$ and $D_2(a)$, the general Green's function is
\begin{align}
\mathrm{GF}(a,b) = \frac{D_1(a) D_2(b) - D_1(b) D_2(a)}{D'_1(b) D_2(b) - D_1(b) D'_2(b)},
\label{eq:GreensFunction}
\end{align}
where $'$ denotes a derivative with respect to $b$.  Given the homogeneous solutions $C_1$ and $C_2 a$, the Green's function for an era of kination is $(a-b)$.  Therefore, the particular solution (PS) is
\begin{align}
\mathrm{PS}(a) = \int_{0}^{a} \mathrm{d}b \,S(k,b) \left(a-b \right).
\label{eq:ParticularFunction}
\end{align}
The particular solution and its derivative equal zero at ${a=0}$.  If we neglect the effects of dark matter annihilations, the adiabatic initial condition for $\delta_\chi$ requires ${\delta_\chi = \Phi_0}$ and ${\delta_\chi'(a) =0}$ at ${a = 0}$, which implies that ${C_2 = 0}$ and ${C_1 = \Phi_0}$.  Combining the homogeneous and particular solution produces the final analytic expression for $\delta_\chi$:
\begin{align}
\delta_\chi = \Phi_0 + \int_{0}^{a} \mathrm{d}b \, S(k,b) \left(a-b \right).
\label{eq:DMTotal}
\end{align}

Using the analytic expression for $\Phi$ given by Eq.~(\ref{eq:PhiKination}), we evaluate the source term and determine the evolution of $\delta_\chi$ well after the mode enters the horizon.  If the integral in Eq.~(\ref{eq:DMTotal}) is evaluated for ${a \gg a_\mathrm{hor}}$, changing the upper bound on the integral from $a$ to $\infty$ does not change the value of the integral because the source term goes to zero at ${a\gg a_\mathrm{hor}}$.  With this approximation, we find that, well after horizon entry, $\delta_\chi$ is equal to the sum of a constant term and a term that grows linearly with the scale factor:
\begin{subequations}
\begin{align}
& \delta_\chi = A + B\, a, \\
& A = \Phi_0 + \int_{0}^{\infty} \mathrm{d}b \, S(k,b) \left(-b \right) = \Phi_0 - \Phi_0 = 0, \\
& B = \int_{0}^{\infty} \mathrm{d}b \, S(k,b) = 2 \, \frac{\Gamma(3/4)}{\Gamma(5/4)}\, \tilde{k}^{1/2} \, \Phi_0 \simeq 2.7\, \tilde{k}^{1/2} \, \Phi_0.
\end{align}
\label{eq:IndividualTermsParticularSolution}%
\end{subequations}
Therefore, if a perturbation mode enters the horizon during kination, the subhorizon evolution of $\delta_\chi$ for ${a \gg a_\mathrm{hor}}$ during kination is
\begin{align}
\delta_\chi = 2.7\, \tilde{k}^{1/2} \Phi_0 \, a = 2.7\,\Phi_0 \left(\frac{a}{a_\mathrm{hor}} \right).
\label{eq:DMKinationDomination}
\end{align}
The last equality follows from the fact that we defined ${\tilde{k} \equiv k/H_1 = a_\mathrm{hor} H(a_\mathrm{hor})/H_1 = a_\mathrm{hor} E(a_\mathrm{hor})}$.  During an era of kination, ${E(a) = a^{-3}}$, and therefore ${a_\mathrm{hor} = \tilde{k}^{-1/2}}$.
\begin{figure*}[t]
 \centering
\begin{minipage}{0.5\textwidth}
\centering
 \resizebox{3.4in}{!}
 {
      \includegraphics{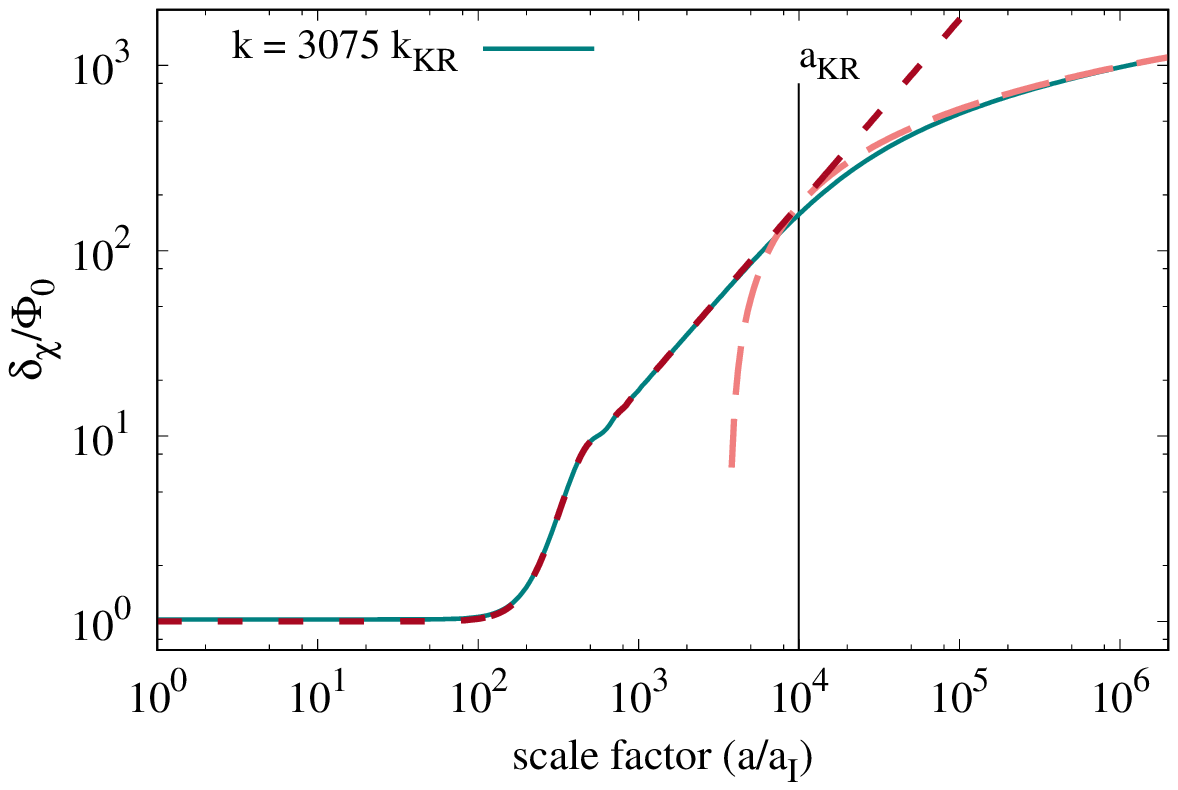}
 }
\end{minipage}%
\begin{minipage}{0.5\textwidth}
\centering
 \resizebox{3.4in}{!}
{
      \includegraphics{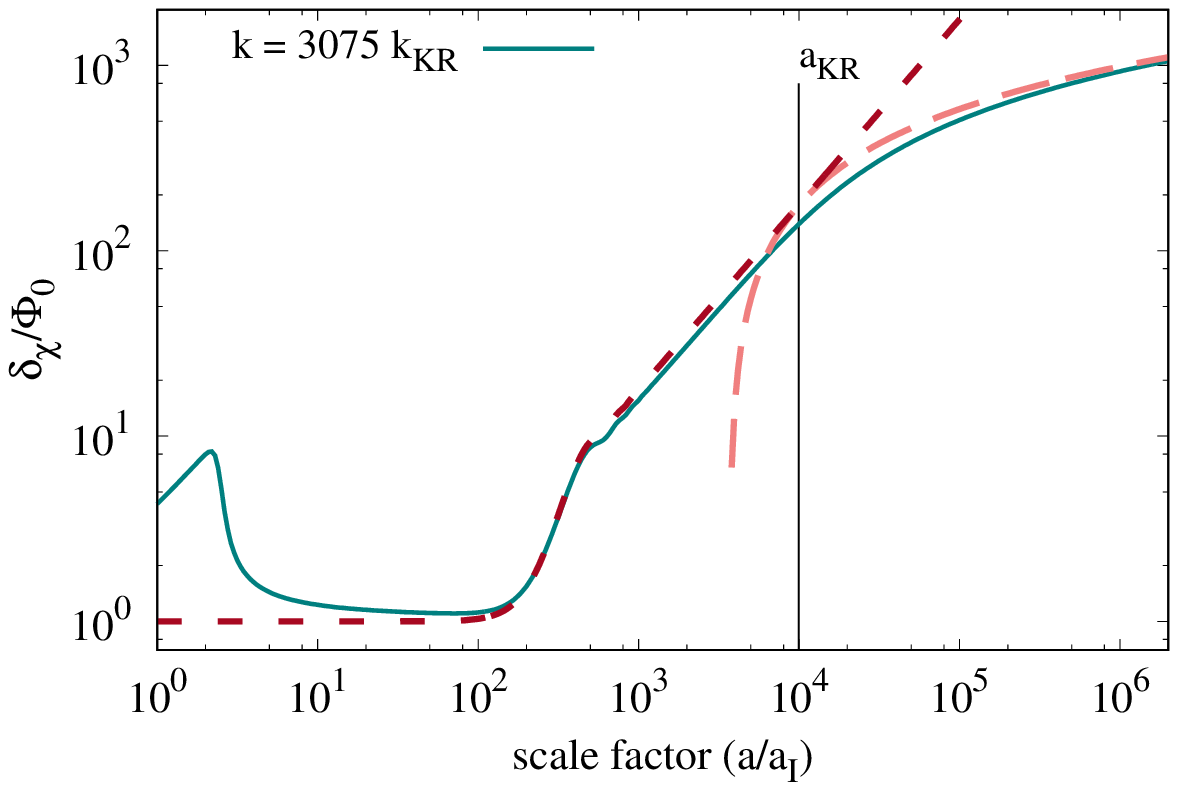}
 }
\end{minipage}%
\caption{Evolution of a dark matter density perturbation that enters the horizon during an era of kination.  The left panel corresponds to scenarios where dark matter freezes in during an era of kination with ${\langle \sigma v \rangle = 1.4 \times 10^{-46} \sigunits}$.  The right panel corresponds to scenarios where dark matter freezes out during an era of kination with ${\langle \sigma v \rangle = 1 \times 10^{-23} \sigunits}$ and freeze-out occurring at ${a_\mathrm{F}=2.5}$.  In both panels, ${m_\chi = 10^5 \, \mathrm{GeV}}$ and ${a_\mathrm{KR} = 10^4}$.   The mode shown has wave number ${k = 3075 \, k_{\mathrm{KR}}}$ and enters the horizon at ${a = 150}$.  The solid curve represents the numerical solution for $\delta_\chi$ using Eq.~(\ref{pertsa}), the short-dashed line represents the analytical expression for $\delta_\chi$ using Eq.~(\ref{eq:DMTotal}), and the long-dashed line corresponds to Eq.~(\ref{eq:DMRadiationDomination}).}
\label{Fig:AnalyticChiEvolution}
\end{figure*}

After the Universe becomes radiation dominated, $\delta_\chi$ grows logarithmically and $a\delta'_\chi$ is constant: ${a\delta'_\chi(a) \simeq a_{\mathrm{KR}} \delta'_{\chi}(a_{\mathrm{KR}})}$.  Solving for the evolution of $\delta_\chi$ after kinaton-radiation equality and connecting it with Eq.~(\ref{eq:DMKinationDomination}) yields
\begin{align}
\delta_\chi = 2.7\, \Phi_0\, \left( \frac{a_{\mathrm{KR}}}{a_\mathrm{hor}} \right) \left[1+\mathrm{ln}\left(\frac{a}{a_{\mathrm{KR}}} \right) \right].
\label{eq:DMRadiationDomination}
\end{align}
Figure \ref{Fig:AnalyticChiEvolution} shows the evolution of $\delta_\chi$ using the numeric solution to Eq.~(\ref{pertsa}) and the analytic expressions given by Eqs.~(\ref{eq:DMTotal}) and (\ref{eq:DMRadiationDomination}) for freeze-in and freeze-out scenarios.  Equation~(\ref{eq:DMTotal}) matches the numeric expression for $\delta_\chi$ to within $5\%$ for ${a < 0.5 \, a_\mathrm{KR}}$ for freeze-in scenarios.  We see in the right panel of Figure \ref{Fig:AnalyticChiEvolution} that the numeric evolution of $\delta_\chi$ does not match the analytic evolution for freeze-out scenarios before horizon entry.  This discrepancy is due to the fact that Eq.~(\ref{eq:DMTotal}) was derived neglecting dark matter annihilations and assuming that ${\delta_\chi = \Phi_0}$ at ${a = 0}$.  For freeze-out scenarios, the condition that ${\delta_\chi = \Phi_0}$ at ${a = 0}$ is not valid, but $\delta_\chi$ does evolve toward $\Phi_0$ after freeze-out, as seen in Figure \ref{Fig:ChiEvolution}.  As $a_\mathrm{hor}$ becomes much larger than $a_\mathrm{F}$, $\delta_\chi$ gets closer to $\Phi_0$ before horizon entry, and Eq.~(\ref{eq:DMTotal}) better matches the numeric solution to Eq.~(\ref{pertsa}).  For example, with ${a_\mathrm{hor}/a_\mathrm{F} = 485}$, Eq.~(\ref{eq:DMTotal}) must be multiplied by a factor of 0.95 to match the numeric evolution of $\delta_\chi$ between ${a_\mathrm{hor} < a < 0.5 \, a_\mathrm{KR}}$.  If ${a_\mathrm{hor}/a_\mathrm{F} = 57}$, the correction factor varies between 0.85 and 0.95 for different $a$ values.  These corrections must also be taken into account when comparing Eq.~(\ref{eq:DMRadiationDomination}) to the numeric evolution of $\delta_\chi$ for freeze-out scenarios.  For freeze-in scenarios, Eq.~(\ref{eq:DMRadiationDomination}) matches the numerics after $a = 10\,a_\mathrm{KR}$ to within $5\%$.

To compare the evolution of $\delta_\chi$ for modes that enter the horizon during an era of kination to those that enter during radiation domination, we repeat our previous analysis for modes that enter the horizon when the dominant component of the Universe has a generic equation-of-state parameter $w$.  Given that the dominant component of the Universe has a constant $w$, the evolution equations for $\delta_\chi$ and $\tilde{\theta}_\chi$ are
\begin{subequations}
\begin{align}
& \delta'_\chi = -\tilde{\theta}_\chi a^{\frac{-3}{2}(\frac{1}{3}-w)} - 3\Phi', \\
& \tilde{\theta}'_\chi = -\frac{\tilde{\theta}_\chi}{a} - \tilde{k}^2 a^{\frac{-3}{2}(\frac{1}{3}-w)} \Phi.
\end{align}
\label{eq:GeneralWPerturbationsDM}%
\end{subequations}
Combining these equations results in a second-order differential equation for $\delta_\chi$:
\begin{align}
\delta^{\prime \prime}_\chi + \frac{3}{2}(1-w)\frac{\delta^\prime_\chi}{a}  = \tilde{k}^2 a^{-3(\frac{1}{3}-w)} \Phi - \frac{9}{2}(1-w)\frac{\Phi^\prime}{a} - 3\Phi^{\prime \prime}.
\label{eq:DMSecondOrderGeneralW}
\end{align}
The homogeneous equation corresponding to Eq.~(\ref{eq:DMSecondOrderGeneralW}) is ${\delta^{\prime \prime}_\chi + \frac{3}{2}(1-w)\frac{\delta^\prime_\chi}{a}=0}$. If $w \neq 1/3$, the homogeneous solution is
\begin{align}
\delta_\chi = C_1 + C_2 \, a^{\frac{3}{2}w-\frac{1}{2}},
\label{eq:DMHomogeneousGeneralW}
\end{align}
whereas if $w=1/3$, the homogeneous solution is
\begin{align}
\delta_\chi = C_1 + C_2 \, \mathrm{ln}(a).
\label{eq:DMHomogeneousRD}
\end{align}
We showed in Figure \ref{Fig:PhiEvolutionAnalyticGenericW} that the evolution of $\Phi$ is qualitatively the same for perturbation modes that enter the horizon when the dominant component of the Universe has $w>0$.  Therefore, the source term will also be qualitatively the same for these scenarios.  Since the integral of the source term is constant at late times, the Green's function produces a similar functional form for the evolution of $\delta_\chi$ compared to the homogeneous solution.  Therefore, if a perturbation mode enters the horizon and ${w > 0}$ and ${w \neq 1/3}$, then the Green's function demands
\begin{align}
\delta_\chi = A + B \, a^{\frac{3}{2}w-\frac{1}{2}},
\label{eq:EvolutionGeneralW}
\end{align}
and, if ${w = 1/3}$, then
\begin{align}
\delta_\chi = A + B \, \mathrm{ln}(a),
\label{eq:EvolutionRD}
\end{align}
where $A$ and $B$ are integration constants.  Overall, the logarithmic growth experienced by subhorizon matter perturbations during radiation domination is a by-product of the homogeneous solution to Eq.~(\ref{eq:DMSecondOrderGeneralW}).  Similarly, it is the homogeneous solution that leads to the linear growth of subhorizon matter perturbations during an era of kination.

The different $\delta_\chi$ growth rates can be attributed to the motion of the dark matter particles.  We saw in Figure \ref{Fig:ChiEvolution} that once a mode enters the horizon, the dark matter density perturbation experiences a kick from the decaying gravitational potential.  This kick causes the dark matter particles to drift toward overdense areas, even after $\Phi \rightarrow 0$.  The comoving displacement of massive particles is
\begin{align}
\vec{s} = \int v \frac{\mathrm{d}t}{a} \propto \int \frac{\mathrm{d}t}{a^2} \propto \int \frac{\mathrm{d}a}{a^3 \, H},
\label{eq:ComovingDisplacement}
\end{align}
where ${v\propto 1/a}$ is the physical particle velocity.  Expressing the Hubble parameter in terms of $w$, Eq.~(\ref{eq:ComovingDisplacement}) implies that ${\vec{s} \propto a^{(3w-1)/2}}$ for ${w > 1/3}$ and ${\vec{s} \propto \mathrm{ln}\,a}$ for ${w = 1/3}$.  To linear order, $\delta$ evolves in the same manner as $\vec{s}$: ${\delta = - \vec{\nabla} \cdot \vec{s}}$.  Therefore, $\delta_\chi$ grows as a direct consequence of the particles' drift toward overdense regions.  If ${0 < w < 1/3}$, $\delta_\chi$ does not grow because the expansion of the comoving grid is faster than the particles' drift velocity.  During radiation domination, the comoving drift velocity decays as $a^{-2}$, which is the same as the expansion of the comoving grid ${(H \propto a^{-2})}$, resulting in logarithmic growth.  During an era of kination, $H(a)$ decreases faster than during radiation domination, so the expansion of the comoving grid slows down faster, thereby allowing a particle with a given drift velocity to cover more comoving space, which results in an enhanced growth rate for $\delta_\chi$.  It is important to note that this mechanism for dark matter perturbation growth is different than that experienced during matter domination.  During matter domination, the gravitational potential is constant in time and the dark matter particles experience a perpetual gravitational force, thereby causing $\delta_\chi$ to grow linearly with the scale factor.

The fact that dark matter particles are drifting toward initially overdense regions does not necessarily mean structure will form during an era of kination.  In other words, it is still uncertain how $\delta_\chi$ will evolve in the non-linear regime.  One possibility is that dark matter particles are moving fast enough that, instead of collapsing and forming structure, they pass by each other and overdense regions becomes underdense.  Collapse may still be possible, however, if local areas of matter domination persist long enough to halt the motion of particles through the overdense region.  Further investigation is required to determine how $\delta_\chi$ evolves in the non-linear regime during an era of kination or radiation domination.  However, we can be certain that modes that remain linear until matter-radiation equality will form halos.  In addition, these halos form earlier than they would if the Universe had been radiation dominated when the relevant scales entered the horizon, due to the enhanced growth of $\delta_\chi$ during kination.

%%%%%%%%%%%%%%%
\section{The Matter Power Spectrum}
\label{sec:MatterPowerSpectrum}
%%%%%%%%%%%%%%%

\begin{figure}
\centering\includegraphics[width=3.4in]{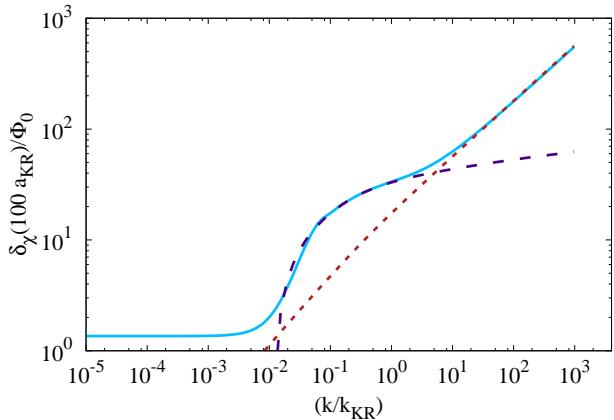}
\caption{The dark matter density perturbation evaluated at ${100 \, a_{\mathrm{KR}}}$ for various $k$ modes.  In this figure, ${m_\chi = 10^5 \, \mathrm{GeV}}$, ${\langle \sigma v \rangle = 1.4 \times 10^{-46} \sigunits}$, and kinaton-radiation equality occurs at ${a_\mathrm{KR} = 10^4}$.  The solid curve represents the numeric evaluation of Eq.~(\ref{pertsa}), while the long-dashed line and short-dashed line represent the analytical evolution via Eqs.~(\ref{eq:TransferFunctionRD}) and (\ref{eq:TransferFunctionKD}).}
\label{Fig:TransferFunction}
\end{figure}

To analyze the matter power spectrum, we evaluate $\delta_\chi$ at a fixed time for various $k$ values.  Figure \ref{Fig:TransferFunction} shows how $\delta_\chi$ changes with $k$ when numerically evaluating Eq.~(\ref{pertsa}) at a fixed value of the scale factor well after kinaton-radiation equality.  The scenario depicted is a freeze-in scenario with ${m_\chi = 10^5 \, \mathrm{GeV}}$ and ${\langle \sigma v \rangle = 1.4\times 10^{-46} \sigunits}$.  The transfer functions for freeze-in and freeze-out scenarios differ by less than $4\%$ for modes with ${a_\mathrm{hor}/a_\mathrm{F} \gtrsim 800}$: these modes are unaffected by ``relentless" annihilations.  In Figure \ref{Fig:TransferFunction}, there are three distinct behaviors.  Modes with ${k \lesssim 0.01 \, k_{\mathrm{KR}}}$ are still outside the horizon at ${a = 100\, a_\mathrm{KR}}$.  To preserve adiabaticity, superhorizon modes evolve and $\delta_\chi$ increases by a factor of 4/3 as the Universe transitions from an era of kination to radiation domination.  Thus, ${\delta_\chi (100 \, a_\mathrm{KR}) = (4/3)\Phi_0}$.

Modes with ${0.01 \lesssim k/k_{\mathrm{KR}} \lesssim 1}$ enter the horizon during radiation domination.  The evolution of $\delta_\chi$ for these modes follows the function
\begin{align}
\delta_\chi(a) = \Phi_\mathrm{p} \left[A \, \mathrm{ln}\left(\frac{B \, a}{a_\mathrm{hor}} \right) \right],
\label{eq:TransferFunctionRD}
\end{align}
with $A = 9.11$, $B = 0.594$ \cite{Hu:1995}.  In this expression, $\Phi_\mathrm{p}$ is defined as the superhorizon gravitational potential during radiation domination.  To determine how $\Phi_\mathrm{p}$ relates to $\Phi_0$, the superhorizon gravitational potential during an era of kination, we evolve a superhorizon mode across $a_\mathrm{KR}$.  As the Universe transitions from an era of kination to radiation domination, a superhorizon mode will evolve to keep the curvature perturbation ${\zeta \equiv \Phi + 2\Phi/(3+3w)}$ constant.  Since  ${w=1}$ during an era of kination, ${\zeta_\mathrm{K} (a < a_\mathrm{KR}) = (4/3)\, \Phi_{0}}$.  During radiation domination ${w = 1/3}$, and ${\zeta_\mathrm{R} (a>a_\mathrm{KR}) = (3/2)\, \Phi_{\mathrm{p}}}$.  Since ${\zeta_\mathrm{K} = \zeta_\mathrm{R}}$ for a superhorizon mode, then ${{\Phi_{\mathrm{p}}} = (8/9)\,\Phi_0}$.  The long-dashed line in Figure \ref{Fig:TransferFunction} corresponds to Eq.~(\ref{eq:TransferFunctionRD}) evaluated at ${a = 100 \, a_{\mathrm{KR}}}$ with ${\Phi_\mathrm{p} = (8/9)\Phi_0}$.  This analytical model matches the numeric solution to Eq.~(\ref{pertsa}) extremely well for modes with ${0.05 \lesssim k/k_{\mathrm{KR}} \lesssim 1}$.  As expected, if a mode enters the horizon during radiation domination, it is unaffected by the preceding era of kination.

Modes with ${k/k_{\mathrm{KR}} \gtrsim 1}$ enter the horizon during an era of kination.  We wish to express the evolution of $\delta_\chi$ for these modes in the same fashion as Eq.~(\ref{eq:TransferFunctionRD}).  From Eq.~(\ref{eq:DMRadiationDomination}), we see that ${A = (9/8)\times 2.7 \, a_\mathrm{KR} \tilde{k}^{1/2}}$ and ${\mathrm{ln}(B) = 1 + \mathrm{ln}(a_\mathrm{hor}/a_\mathrm{KR})}$.  Since Figure \ref{Fig:TransferFunction} is in terms of $k/k_\mathrm{KR}$, we similarly need to express $A$ and $B$ in terms of $k/k_\mathrm{KR}$.  Using the fact that ${a_\mathrm{hor} = \tilde{k}^{-1/2}}$ during an era of kination and that ${a_\mathrm{KR} = \tilde{k}_\mathrm{KR}^{-1/2} 2^{1/4}}$, we determine that for modes with ${k/k_{\mathrm{KR}} \gtrsim 1}$,
\begin{align}
\delta_\chi(a) = 2.7 \, \Phi_0 \left(\frac{ k \sqrt{2}}{k_\mathrm{KR}} \right)^{1/2} \mathrm{ln}\left[e\left(\frac{k_\mathrm{KR}}{k \sqrt{2}}\right)^{1/2} \frac{a}{a_\mathrm{hor}} \right].
\label{eq:TransferFunctionKD}
\end{align}
The short-dashed line in Figure \ref{Fig:TransferFunction} correspond to Eq.~(\ref{eq:TransferFunctionKD}), where $a$ is evaluated at ${100 \, a_{\mathrm{KR}}}$.  This analytical model matches the numeric solution very well for modes with ${k/k_{\mathrm{KR}} \gtrsim 100}$.  We also found a fitting function that not only smoothly connects Eqs.~(\ref{eq:TransferFunctionRD}) and (\ref{eq:TransferFunctionKD}), but also matches the numeric solution of $\delta_\chi$ to within $5\%$ for modes with ${0.05 < k/k_{\mathrm{KR}} < 1000}$:
\begin{align}
\delta_\chi(a) &= \frac{8}{9} \Phi_0 \left[A \, \mathrm{ln}\left(\frac{B \, a}{a_\mathrm{hor}} \right) \right] \nonumber \\
A &= 2.29 \left[0.11\times 9.11^{2.64} + 2.9\left(\frac{k}{k_\mathrm{KR}}\right)^{1.32} \right]^{.38} \nonumber \\
B &= \left[0.594^{-6.59} + e^{-6.59} \left(\frac{k}{k_\mathrm{KR}} \right)^{3.29} \right]^{-.15}.
\label{eq:FittedFunction}
\end{align}

From Eq.~(\ref{eq:TransferFunctionKD}), it is clear that if a mode enters the horizon during an era of kination, ${\delta_\chi/\Phi_0 \propto k^{1/2}}$.  In contrast, it has been shown in Refs.~\cite{Erickcek:2011,Erickcek:2015} that if a mode enters the horizon during an EMDE, ${\delta_\chi/\Phi_0 \propto k^2}$.  Both of these scalings are consistent with linear growth prior to the onset of radiation domination.  Since a mode enters the horizon at ${k = aH}$, during an era of kination ${a_\mathrm{hor} \propto k^{-1/2}}$, and during an EMDE ${a_\mathrm{hor} \propto k^{-2}}$ \cite{Erickcek:2011,Erickcek:2015}.  Therefore, even though ${\delta_\chi \propto a/a_\mathrm{hor}}$ during both eras, $\delta_\chi$ scales differently with $k$.

With this scaling, we can determine how the matter power spectrum scales with $k$ for modes with ${k> k_\mathrm{KR}}$. The power spectrum of density perturbations is ${P_\delta = P_\Phi (\delta/\Phi_0)^2}$, where ${P_\Phi \propto k^{n_s-4}}$ is the power spectrum of curvature fluctuations and $n_s$ is the scalar spectral index.  Since ${\delta_\chi/\Phi_0 \propto k^{1/2}}$ for modes that enter the horizon during an era of kination, ${P_\delta \propto k^{n_s-3}}$ for ${k > k_\mathrm{KR}}$.  In comparison, ${P_\delta \propto k^{n_s}}$ for modes with ${k<k_\mathrm{eq}}$ and ${P_\delta \propto k^{n_s-4} \, \mathrm{ln}^2k}$ for modes with ${k_\mathrm{eq} < k <k_\mathrm{KR}}$, where $k_\mathrm{eq}$ is the wave number of the mode that enters the horizon at matter-radiation equality.  Thus, there is a small-scale enhancement to the matter power spectrum due to an era of kination compared to modes that enter the horizon during radiation domination.  In addition, the matter power spectrum is shallower on small scales for modes that enter the horizon during kination compared to modes that enter the horizon during an EMDE, implying that collapse at a given scale will happen later following an era of kination compared to following an EMDE.

%%%%%%%%%%%%%%%
\section{Conclusion}
\label{sec:Conclusion}
%%%%%%%%%%%%%%%

We determined how the gravitational potential $\Phi$ and the dark matter density perturbations $\delta_\chi$ evolve for modes that enter the horizon during an era of kination.  In addition to numerically solving for the evolution of $\Phi$ and $\delta_\chi$ during kination, we also derived analytic expressions for their evolution during eras in which the dominant component of the Universe has a general equation-of-state parameter $w$.  We determined that the gravitational potential vanishes upon horizon entry if the dominant energy component has ${w > 0}$.  In addition, if ${w > 1/3}$, then ${\delta_\chi \propto a^{3w/2 - 1/2}}$.  Consequently, if a perturbation mode enters the horizon during an era of kination $(w = 1)$, $\delta_\chi$ grows linearly with the scale factor.  This linear growth is due to the drift of the dark matter toward initially overdense regions.  The comoving displacement of massive particles under pure drift is ${\vec{s} \propto a^{(3w-1)/2}}$ for ${w > 1/3}$.  If ${0<w<1/3}$, the expansion rate of the Universe is greater than the particles' drift velocity and ${\vec{s}}$ cannot grow, thereby not permitting perturbation growth.  For modes that enter the horizon during an era of kination, the linear growth of $\delta_\chi$ during kination implies that ${\delta_\chi/\Phi_0 \propto k^{1/2}}$.  Therefore, for modes that enter the horizon during kination, the matter power spectrum ${P_\delta \propto k^{n_s-3}}$.  In comparison, for modes that enter the horizon during radiation domination ${P_\delta \propto k^{n_s-4} \, \mathrm{ln}^2k}$ and there is a small-scale enhancement to the matter power spectrum due to an era of kination compared to modes that enter the horizon during radiation domination.

The linear growth experienced by $\delta_\chi$ during kination will lead to enhanced small-scale structure formation following kination.  The presence of small-scale structure effectively increases the dark matter annihilation rate.  If a perturbation mode enters the horizon during an EMDE, $\delta_\chi$ grows linearly with the scale factor and the dark matter annihilation rate is enhanced by several orders of magnitude, depending on the formation time of the microhalos \cite{Erickcek:2011,Erickcek:2015}.  However, the boost to the dark matter annihilation rate is limited to $10^6$ if halos only form after $z \sim 500$ \cite{Erickcek:2011,Erickcek:2015}.  In Ref.~\cite{Redmond:2017} we found that in order to produce the observed dark matter relic abundance, freeze-in scenarios during an era of kination require $\langle \sigma v \rangle$ be between ${10^{-50} \sigunits}$ and ${10^{-37} \sigunits}$ for ${10^{-2} \, \mathrm{GeV} \lesssim m_\chi< 10^5 \, \mathrm{GeV}}$.  The strongest observational bound on $\langle \sigma v \rangle$ is ${\langle \sigma v \rangle < 10^{-27} \sigunits}$, and that bound is applicable to ${1 \, \mathrm{GeV} \lesssim m_\chi \lesssim 10 \, \mathrm{GeV}}$ \cite{Fermi:Constraints}.  Therefore, even with a boost factor of $10^6$, the dark matter annihilation signal for freeze-in scenarios during an era of kination is still not observable.

For freeze-out scenarios during an era of kination, the $\langle \sigma v \rangle$ values that would produce the observed dark matter relic abundance are nearly ruled out by Fermi-LAT\cite{Fermi:Constraints} and H.E.S.S.\cite{HESS:Constraints} observations.  A modest boost factor of 10 would completely rule out freeze-out scenarios during an era of kination if dark matter annihilates via the $b\overline{b}$, $\tau^+ \tau^-$, or $W^+W^-$ annihilation channels.  It remains to be determined if the growth of perturbations during kination is capable of generating this enhancement.  Although an EMDE can easily lead to boost factors of order 1000 or more, the linear growth of perturbations during kination leads to a shallower enhancement to the small-scale power spectrum: ${P(k) \propto k^{n_s-3}}$ instead of ${P(k) \propto k^{n_s}}$.  As a result, the formation of microhalos will be delayed relative to EMDE cosmologies because a given value of ${k/k_\mathrm{KR}}$ will go nonlinear far later than the same value of ${k/k_\mathrm{RH}}$, where $k_\mathrm{RH}$ is the scale that enters the horizon at the end of an EMDE.  Equivalently, larger values of ${k/k_\mathrm{KR}}$ are required to form microhalos at a given redshift, which increases the likelihood that these scales will be suppressed by the free-streaming of the dark matter particles \cite{Visinelli:2015}.  An analysis of the microhalo abundance predicted by Press-Schechter theory \cite{Press:1973} following the same procedure as Refs.\cite{Erickcek:2011, Erickcek:2015} would provide an estimate of the boost factor and determine if freeze-out scenarios during kination are compatible with current observational bounds on dark matter annihilation.

\section*{Acknowledgments}
We thank M. Sten Delos and Chris Hirata for insightful and helpful discussions.  This work was supported by NSF Grant No. PHY-1417446.

%%%%%%%%%%%%%%%%%%%%%%%%%%%
\appendix
\section{Derivation of the Perturbation Equations}
\label{sec:perts}
%%%%%%%%%%%%%%%%%%%%%%%%%%%

The perturbation evolution equations are derived by perturbing the covariant form of the energy-transfer equations given in Eq.~(\ref{eq:Boltz}).  We follow the same approach as that outlined in Refs.~\cite{Erickcek:2011,Barenboim:2013,Fan:2014,Erickcek:2015}.  The kinaton, dark matter, and radiation all behave as perfect fluids with energy momentum tensors
\begin{align}
T^{\mu\nu} = pg^{\mu\nu} + (\rho + p)u^\mu u^\nu,
\label{eq:EnergyMomentumTensor}
\end{align}
where $\rho$ is the fluid's energy density, $p$ is its pressure, and ${u^\mu \equiv \mathrm{d}x^\mu / \mathrm{d}\lambda}$ is its four-velocity.  The kinaton has ${p = \rho}$, the radiation has ${p = \rho/3}$, and the dark matter is a pressureless fluid.  Equation~(\ref{eq:Boltz}) implies that the three fluids exchange energy.  This energy exchange is expressed covariantly as
\begin{align}
\nabla_\mu \left( ^{(i)}T^\mu_\nu  \right) = Q^{(i)}_\nu,
\label{eq:CovariantEnergyExchange}
\end{align}
where $i$ represents the individual fluids.  In the absence of spatial variations,
\begin{subequations}
\begin{align}
& \nabla_\mu \left( ^{(i)}T^\mu_0  \right) = -\dot{\rho}_i - 3H(\rho_i + p_i), \\
& \nabla_\mu \left( ^{(i)}T^\mu_j  \right) = 0,
\end{align}
\label{eq:CovariantVariation}%
\end{subequations}
where a dot represents differentiation with respect to proper time.  Using Eq.~(\ref{eq:CovariantVariation}) and Eq.~(\ref{eq:Boltz}), the covariant energy exchange for this three-fluid model is
\begin{subequations}
\begin{align}
& Q^{(\phi)}_\nu = 0, \\
& Q^{(\chi)}_\nu = -L_\nu, \\
& Q^{(r)}_\nu = L_\nu,
\end{align}
\label{eq:Qs}%
\end{subequations}
where
\begin{align}
L_\nu = \frac{\langle \sigma v \rangle}{m_\chi}\left(\rho_\chi^2 \, u_\nu^{(\chi)} - \rho_{\chi,\mathrm{eq}}^2 \, u_\nu^{(r)} \right).
\label{eq:Lnu}
\end{align}
Equation~(\ref{eq:Lnu}) is different than the definition of $L_\nu$ in Ref.~\cite{Erickcek:2015}.  We have corrected the expression for $L_\nu$ to account for the fact that, while in thermal equilibrium, the dark matter is pair-produced with the same velocity as the radiation. This change introduces coupling terms between $\theta_\chi$ and $\theta_r$ that are only relevant while pair production is important.

Next, we evaluate Eq.~(\ref{eq:CovariantEnergyExchange}) using the perturbed Friedmann-Robertson-Walker (FRW) metric
\begin{align}
\mathrm{d}s^2 = -\left(1+2\Psi \right)\mathrm{d}t^2 + a^2(t)\delta_{ij}\left(1+2\Phi \right)\mathrm{d}x^i \mathrm{d}x^j.
\label{eq:PerturbedFRW}
\end{align}
Perturbations are introduced into the density of each fluid with ${\rho_i(t,\vec{x}) = \rho_i^0(t)\left[1+\delta_i(t,\vec{x}) \right]}$, where $\rho_i^0(t)$ is the background energy density of each fluid and ${\delta_i(t,\vec{x}) \equiv \delta\rho_i / \rho_i^0}$ is the density perturbation of each fluid.  In addition, perturbations are introduced into the four-velocity of each fluid: ${u_0 = -(1+\Psi)}$ and ${u_{j(i)} = a^2 \delta_{kj}v^k_{(i)}}$, where ${v^k_{(i)} \equiv \mathrm{d}x^k/\mathrm{d}t}$ is the peculiar velocity of the $i$th fluid.  The combination of perturbations to the metric, energy density, and four-velocity introduce perturbations to the energy exchange variables $Q_\nu$ and $L_\nu$: the first-order components are
\begin{align}
Q^{(\phi),(1)}_0 = Q^{(\phi),(1)}_j = 0,
\label{eq:PerturbedQ}
\end{align}
\begin{align}
L^{(1)}_0 = -\frac{\langle \sigma v \rangle}{m_\chi} \left[ \left(\rho^0_\chi\right)^2 \left( 2\delta_\chi + \Psi \right) - \left(\rho^0_{\chi,\mathrm{eq}}\right)^2 \left( 2\delta_{\chi,\mathrm{eq}}+ \Psi \right) \right],
\label{eq:PerturbedLo}
\end{align}
\begin{align}
L^{(1)}_j = \frac{\langle \sigma v \rangle}{m_\chi} a^2 \left[\left(\rho^0_\chi\right)^2 \delta_{ij} v^i_{(\chi)} - \left(\rho^0_{\chi,\mathrm{eq}}\right)^2 \delta_{ij} v^i_{(r)} \right],
\label{eq:PerturbedLj}
\end{align}
where $\delta_{\chi,\mathrm{eq}}$ is the dark matter equilibrium density perturbation.  We see that both the zeroth- and first-order components of $Q^{(\phi)}_0$ and $Q^{(\phi)}_j$ are zero, whereas $L_\nu$ contains both a zeroth- and first-order component.

Taking into account first-order perturbations, the ${\mu = 0}$ component of Eq.~(\ref{eq:CovariantEnergyExchange}) requires that each fluid obey the equation
\begin{align}
\frac{\mathrm{d}\delta_i}{\mathrm{d}t} + \left(1+w_i \right) \frac{\theta_i}{a}\, +& \, 3\left(1+w_i \right)\frac{\mathrm{d}\Phi}{\mathrm{d}t} = \nonumber \\
&\frac{1}{\rho^0_i}\left(Q^{(i),(0)}_0 \delta_i - Q^{(i),(1)}_0 \right),
\label{eq:Delta}
\end{align}
where $w_i$ is the equation of state parameter for a given fluid, $\theta_i \equiv a\partial_j v^j$ is the divergence of the fluid's physical velocity, and $Q^{(i),(0)}_0$ and $Q^{(i),(1)}_0$ are the zeroth-order and first-order components of $Q_0^{(i)}$ for each fluid.  The divergence of the spatial component of Eq.~(\ref{eq:CovariantEnergyExchange}) requires that each fluid obey the equation
\begin{align}
\frac{\mathrm{d}\theta_i}{\mathrm{d}t} + \left(1-3w_i \right)&H\theta_i + \frac{\nabla^2 \Phi}{a} + \frac{w_i}{1+w_i}\frac{\nabla^2 \delta_i}{a} = \nonumber \\
&\frac{1}{\rho^0_i}\left(\frac{\partial_i Q_i}{a\left(1+w_i\right)} + Q^{(i),(0)}_0 \theta_i \right).
\label{eq:ThetaEvolution}
\end{align}

Applying Eqs.~(\ref{eq:Delta}) and (\ref{eq:ThetaEvolution}) to the kinaton (${w_k = 1}$), dark matter (${w_\chi = 0}$), and radiation (${w_r = 1/3}$) yields the following perturbation equations
\begin{widetext}
\begin{subequations}
\begin{align}
&\frac{\mathrm{d}\delta_\phi}{\mathrm{d}t}+\frac{2\theta_\phi}{a} + 6\frac{\mathrm{d}\Phi}{\mathrm{d}t} = 0,\\
& \frac{\mathrm{d} \theta_\phi}{\mathrm{d}t} -2H\theta_\phi +\frac{\nabla^2 \Psi}{a} + \frac{1}{2}\frac{\nabla^2 \delta_\phi}{a}=0,\\
& \frac{\mathrm{d}\delta_\chi}{\mathrm{d}t}+\frac{\theta_\chi}{a} + 3\frac{\mathrm{d}\Phi}{\mathrm{d}t} = \frac{\langle \sigma v \rangle}{m_\chi \rho^0_\chi} \left[-\Psi\left\{\left(\rho^0_\chi\right)^2 - \left(\rho^0_{\chi,\mathrm{eq}}\right)^2 \right\} - \left(\rho^0_\chi\right)^2 \delta_\chi + \left(\rho^0_{\chi,\mathrm{eq}}\right)^2 \left(2\delta_{\chi,\mathrm{eq}}-\delta_\chi \right) \right],\\
& \frac{\mathrm{d} \theta_\chi}{\mathrm{d}t} +H\theta_\chi +\frac{\nabla^2 \Psi}{a} = \frac{\langle \sigma v \rangle \left( \rho_{\chi,\mathrm{eq}}^0\right)^2}{m_\chi \rho_\chi^0} \left(\theta_r - \theta_\chi \right) ,\\
& \frac{\mathrm{d}\delta_r}{\mathrm{d}t}+\frac{4}{3}\frac{\theta_r}{a} + 4\frac{\mathrm{d}\Phi}{\mathrm{d}t} = \frac{\langle \sigma v \rangle}{m_\chi \rho^0_r} \left[\Psi\left\{\left(\rho^0_\chi\right)^2 - \left(\rho^0_{\chi,\mathrm{eq}}\right)^2 \right\} + \left(\rho^0_\chi\right)^2 \left( 2\delta_\chi - \delta_r \right) - \left(\rho^0_{\chi,\mathrm{eq}}\right)^2 \left(2\delta_{\chi,\mathrm{eq}}-\delta_r \right) \right],\\
&\frac{\mathrm{d} \theta_r}{\mathrm{d}t} +\frac{\nabla^2 \Psi}{a} + \frac{1}{4}\frac{\nabla^2 \delta_r}{a}= \frac{\langle \sigma v \rangle}{m_\chi \rho^0_r} \left( \left(\rho_\chi^0\right)^2 \left\{ \frac{3}{4}\theta_\chi - \theta_r \right\} + \frac{1}{4} \left(\rho_{\chi,\mathrm{eq}}^0\right)^2 \theta_r   \right).
\end{align}
\label{perts}%
\end{subequations}
\end{widetext}
The perturbed time-time component of the Einstein equation yields
\begin{align}
\frac{\nabla^2 \Phi}{a^2} + 3H\left(H \Psi - \frac{\mathrm{d}\Phi}{\mathrm{d}t} \right) = -4\pi G \left(\rho_\phi^0 \delta_\phi + \rho_\chi^0 \delta_\chi + \rho_r^0 \delta_r \right).
\label{eq:PerturbedEinsteinTime}
\end{align}
Equation~(\ref{perts}) assumes that the scalar field does not interact with either the dark matter or radiation and also assumes that the dark matter is created solely from thermal production.

%%%%%%%%%%%%%%%
\section{Initial Conditions}
\label{sec:InitialConditions}
%%%%%%%%%%%%%%%

The evolution of density and velocity perturbations for a single plane-wave perturbation mode with wave number $k$ are obtained by numerically integrating Eq.~(\ref{pertsa}) from $a=1$ to some scale factor value well after kinaton-radiation equality.  The integration begins when the mode is well outside the horizon ($k \ll aH$).  To solve for the perturbation initial conditions, we simplify Eq.~(\ref{pertsa}) using the fact that at early times the Universe was dominated by the kinaton, so ${E(a) \approx a^{-3}}$ and ${\tilde{\rho}_\phi \approx a^{-6}}$.

We first solve for the evolution of the kinaton perturbations and the gravitational potential during an era of kination for superhorizon modes.  Simplifying Eqs.~(\ref{pertsa}a), (\ref{pertsa}b), and (\ref{eq:PerturbedEinstein}) yields:
\begin{subequations}
\begin{align}
&\delta'_\phi + 2\,a\, \tilde{\theta}_\phi + 6\Phi' = 0,\\
& \tilde{\theta}'_\phi - \frac{2}{a} \tilde{\theta}_\phi + \tilde{k}^2 a \, \Phi - \frac{1}{2} \, \tilde{k}^2 \, a \, \delta_\phi =0, \\
&\tilde{k}^2 \, \Phi + 3 a^{-4} \left(\Phi' \, a + \Phi  \right) = \frac{3}{2} a^{-4} \, \delta_\phi.
\end{align}
\label{eq:KinationSimplified}%
\end{subequations}
One would initially suspect that given these three equations with three unknowns, we would have a fully defined set of differential equations.  Yet in solving these equations, we discover that there is an ambiguity in the solution to $\tilde{\theta}_\phi$ such that these three equations do not fully define the evolution of $\tilde{\theta}_\phi$ for superhorizon modes.  Equation~(\ref{eq:KinationSimplified}a) corresponds to ${\nabla_\mu T^\mu_0 = 0}$ and Eq.~(\ref{eq:KinationSimplified}b) corresponds to ${\nabla_\mu T^\mu_i = 0}$.  If Eqs.~(\ref{eq:KinationSimplified}) formed a complete set, then they would be able to produce an algebraic expression for $G_i^0$ using the Bianchi identity, which states that ${\nabla_\mu G^\mu_\nu = 0}$.  Yet the Bianchi identity does not provide an algebraic expression for $G_i^0$ and there remains an undermined initial condition.  The time-space component of the perturbed Einstein equation contains additional information regarding our system of equations:
\begin{align}
& \tilde{k}^2 \left(\Phi' \, a + \Phi  \right) = -3 \, a^{-2} \, \tilde{\theta}_\phi.
\label{eq:PerturbedEinsteinTimeSpace}
\end{align}
Equations (\ref{eq:KinationSimplified}) and~(\ref{eq:PerturbedEinsteinTimeSpace}) form a complete set of differential equations and initial conditions.  Utilizing Eqs.~(\ref{eq:KinationSimplified}) and~(\ref{eq:PerturbedEinsteinTimeSpace}), we solve for the evolution of the gravitational potential and the kinaton perturbations for superhorizon modes as an expansion in $\tilde{k}^2$:
\begin{subequations}
\begin{align}
& \Phi = \Phi_0 - \frac{1}{32} \tilde{k}^2 \Phi_0 a^4 + \mathcal{O}(\tilde{k}^4), \\
& \delta_\phi = 2\Phi_0 + \frac{17}{48} \tilde{k}^2 \Phi_0 a^4 + \mathcal{O}(\tilde{k}^4), \\
& \tilde{\theta}_\phi = - \frac{1}{3} \tilde{k}^2 \Phi_0 a^2 + \mathcal{O}(\tilde{k}^4).
\end{align}
\label{eq:InitialConditionsKinatonandPhi}%
\end{subequations}

Since the number of relativistic particles created or destroyed from dark matter annihilations is not sufficient to influence the evolution of $\rho_r$, the interaction between dark matter and radiation will not influence the evolution of radiation perturbations.  Evaluating Eqs.~(\ref{pertsa}e) and~(\ref{pertsa}f) in the superhorizon limit, while ignoring the effects of dark matter annihilations, results in
\begin{subequations}
\begin{align}
& \delta_r = \frac{4}{3}\Phi_0 + \frac{17}{72} \tilde{k}^2 \Phi_0 a^4 + \mathcal{O}(\tilde{k}^4), \\
& \tilde{\theta}_r = - \frac{1}{3} \tilde{k}^2 \Phi_0 a^2 + \mathcal{O}(\tilde{k}^4).
\end{align}
\label{eq:RadiationInitialConditions}%
\end{subequations}

The initial condition for $\delta_\chi$ is chosen to ensure that the perturbations are adiabatic at superhorizon scales.  For freeze-out scenarios, while the dark matter is in thermal equilibrium, the terms on the right-hand side of Eq.~(\ref{pertsa}c) are much larger than the terms on the left-hand side.  To make the terms on the right-hand side vanish while ${\rho_\chi = \rho_{\chi,\mathrm{eq}}}$,
\begin{align}
\delta_\chi = \delta_{\chi,\mathrm{eq}}.
\label{eq:InitialConditionDMEq}
\end{align}
Equation~(\ref{eq:InitialConditionDMEq}) maintains adiabaticity while the dark matter is in thermal equilibrium; if ${\rho_\chi = \rho_{\chi,\mathrm{eq}}}$, then Eq.~(\ref{eq:InitialConditionDMEq}) makes ${\delta_i(\rho_i/\rho^\prime_i)}$ the same for all fluids.

For freeze-in scenarios, the initial condition for $\delta_\chi$ is more difficult to determine from the perturbation equations.  We therefore choose the freeze-in initial condition for $\delta_\chi$ to ensure that ${\delta_i(\rho_i/\rho^\prime_i)}$ is the same for all fluids.  Equations ~(\ref{eq:InitialConditionsKinatonandPhi}b) and (\ref{eq:RadiationInitialConditions}a) already imply that the perturbations to the kinaton and radiation are adiabatic.  To solve for the initial condition for $\delta_\chi$ we set ${\delta_\chi(\rho_\chi/\rho^\prime_\chi) = \delta_\phi(\rho_\phi/\rho^\prime_\phi)}$.  Since ${\rho_\phi \propto a^{-6}}$,
\begin{align}
\delta_\chi = -\frac{1}{6} \delta_\phi \frac{a \, \rho'_\chi(a)}{\rho_\chi(a)}.
\label{eq:InitialConditionDM}
\end{align}
Finally, since adiabatic perturbations require $\theta$ to be the same for all fluids \cite{Weinberg:2003},
\begin{align}
\tilde{\theta}_\chi = - \frac{1}{3} \tilde{k}^2 \Phi_0 a^2 + \mathcal{O}(\tilde{k}^4)
\label{eq:InitialConditionDMTheta}
\end{align}
for superhorizon modes.

\end{document}